\documentclass[10pt,letterpaper]{article}
\usepackage{}
\usepackage{opex3_Bian}

\usepackage{bbm}
\usepackage{mathrsfs}
\usepackage{amsfonts}
\usepackage{amssymb}
\usepackage{graphicx}
\usepackage{cite}
\usepackage[linesnumbered,ruled,lined]{algorithm2e}
\usepackage{array}
\usepackage{subfigure}
\usepackage{multirow}
\usepackage{color}
\usepackage{amsmath}
\usepackage{bm}
\usepackage{upgreek}
\usepackage{float}
\newcommand{\format}[1]{\text{\bf #1}}

\usepackage{fmtcount}
\usepackage{booktabs,threeparttable}
\usepackage[colorlinks,
            linkcolor=black,
            anchorcolor=black,
            citecolor=black,
            urlcolor = blue
            ]{hyperref}

\begin{document}

\title{Fourier ptychographic reconstruction using Wirtinger flow optimization}

\author{Liheng Bian,$^{1}$ Jinli Suo,$^{1}$ Guoan Zheng,$^{2,3}$ KaiKai Guo,$^2$ Feng Chen,$^{1}$ and Qionghai Dai$^{1,*}$}

\address{
$^1$Department of Automation, Tsinghua University, China\\
$^2$Biomedical Engineering, University of Connecticut, USA \\
$^3$Electrical and Computer Engineering, University of Connecticut, USA \\
}

\email{$^*$qionghaidai@tsinghua.edu.cn} 


\begin{abstract}
Recently Fourier Ptychography (FP) has attracted great attention, due to its marked effectiveness in leveraging snapshot numbers for spatial resolution in  large field-of-view imaging. To acquire high signal-to-noise-ratio (SNR) images under angularly varying illuminations for subsequent reconstruction, FP requires long exposure time, which largely limits its practical applications.
In this paper, based on the recently reported Wirtinger flow algorithm, we propose an iterative optimization framework incorporating phase retrieval and noise relaxation together, to realize FP reconstruction using low SNR images captured under short exposure time.
Experiments on both synthetic and real captured data validate the effectiveness of the proposed reconstruction method.
Specifically, the proposed technique could save $\sim 80\%$ exposure time to achieve similar retrieval accuracy compared to the conventional FP. Besides, we have released our source code for non-commercial use.
\end{abstract}


\bibliographystyle{osajnl_suo}
\bibliography{WirtingerFlowForFPM}

\section{Introduction}\label{sec:Introduction}

Fourier Ptychography (FP) is a newly reported technique for large field-of-view (FOV) and high-resolution (HR) imaging \cite{FPM_Nature,FPM_IEEE,FPM_Optics}. This technique sequentially captures a set of low-resolution (LR) images describing different spatial spectrum bands of the sample, and then stitches these spectrum bands together in the Fourier domain to reconstruct the entire HR spatial spectrum, including both amplitudes and phases. This HR spectrum can be transformed to the spatial domain to recover the HR image of the sample. Mathematically, FP reconstruction could be treated as a typical phase retrieval problem, which needs to recover a plural function given the magnitude measurements of its Fourier transform. Specifically, we only obtain the magnitudes of images corresponding to the sub-bands of the HR spatial spectrum, and intend to retrieve the plural HR spectrum. So far, all the FP applications \cite{FPM_Nature, FPM_Quantitative,FPM_Fluorescence,FPM_Cellphone} utilize the alternating projection (AP) algorithm \cite{Phase_Comparison,Phase_Fienup_2}, a widely used method for phase retrieval, to implement the reconstruction process.

Recently, FP technique has been successfully applied to microscopic imaging and brings Fourier ptychography microscopy (FPM) \cite{FPM_Nature}. FPM assumes plane-wave illuminations from the LED array. Therefore, by sequentially lightening LEDs at different positions in the illumination plane, according to \cite{Optics}, we can obtain different shifted versions of the sample's spatial spectrum. Since the whole light field is filtered by the microscope's objective lens, the changing of incident angles results in a set of LR images carrying different sub-bands of the entire spectrum. Finally, by utilizing the FP reconstruction technique (the AP algorithm), FPM can achieve large FOV and HR microscopic imaging. As stated in \cite{FPM_Nature}, the synthetic NA of the FPM setup is $\sim$0.5, and the FOV could reach $\sim$120 mm$^2$, which greatly improves the throughput performance of existing microscopes.

However, the AP algorithm utilized in the reconstruction process is sensitive to the input noise \cite{Phase_Review}, 
and thus long exposure time is required for capturing high signal-to-noise-ratio (SNR) inputs. As reported in \cite{FPM_Nature}, the FPM setup needs around 3 minutes to acquire high SNR images under 137 angularly varying illuminations, for subsequent reconstruction of a gigapixel grayscale image. This largely limits the practical applications of FPM and other FP applications. To shorten the acquisition time, this paper proposes a new phase retrieval method for FP reconstruction, which is able to deal with low SNR input images.

As for existing phase retrieval methods, typically they can be classified into two categories, namely alternating projection algorithms and semi-definite programming (SDP) based algorithms \cite{Phase_Review}. The former kind alternately operates in spatial space and Fourier space, imposing corresponding spatial-plane and Fourier-plane constraints to the retrieved plural function. These constraints include consistency with measured magnitudes \cite{Phase_GS}, magnitudes' non-negativity \cite{Phase_Fienup_1}, signal support constraint \cite{Phase_Comparison}, and so on. This kind of methods are efficient but at the risk of non-convergence and reaching to a local optimum \cite{Phase_Review}. The latter kind of approaches rely on the observation that the quadratic equations in the phase retrieval problem can be rewritten as linear equations in a higher dimensional space \cite{SDP}. Typical SDP algorithms include PhaseLift \cite{Phase_Lift_1,Phase_Lift_2} and PhaseCut \cite{Phase_Cut}, and PhaseLift has been successfully applied to the common phase retrieval task from captured coded diffraction patterns \cite{Phase_Lift_App}. Such kind of methods could converge to a global optimum by a series of convex relaxations. However, they require matrix lifting to work in a higher space, and thus causes high computation cost, which makes them less competitive.

Recently, based on Wirtinger derivatives \cite{Wirtinger_Book_1,Wirtinger_Book_2}, Candes et al. \cite{Phase_Wirtinger} develop a non-convex formulation of the phase retrieval problem, and utilize the gradient descent scheme to derive a computation-cost-saving solution, termed Wirtinger flow (WF) algorithm. As stated in \cite{Phase_Wirtinger}, although the quadratic model is non-convex, Wirtinger flow algorithm can rigorously retrieve exact phase information from a nearly minimal number of random measurements, by starting with a relatively accurate initialization. Besides, the algorithm converges at a geometric rate in the case of random Gaussian sampling mode, and at a linear rate in the case of coded diffraction pattern mode. In this paper, we apply the Wirtinger flow scheme to FP, and further introduce a noise relaxation constraint for a new FP reconstruction framework. The proposed framework is termed WFP (Wirtinger flow optimization for Fourier Ptychographic).
The advantages of the proposed WFP are threefold:
\begin{itemize}
  \item Compared to the existing AP algorithm for FP, WFP can better handle the detector noise, and thus largely reduce the requisite long exposure time;
  \item Compared to the SDP based algorithms such as PhaseLift and PhaseCut, WFP doesn't need matrix lifting and largely decreases the computation cost;
  \item WFP is a general optimization framework being able to incorporate various  priors and constraints, and hence can be extended to save costs and increase the retrieval accuracy further.
\end{itemize}

The remainder of this paper is organized as follows: modeling and derivation of the optimization algorithm are explained in Sec.~\ref{sec:Method}. Then, we conduct a series of experiments on both synthetic and real captured data to validate the proposed approach in Sec.~\ref{sec:Experiments}. Finally, we conclude this paper with some summaries and discussions in Sec.~\ref{sec:Conclusions}.

\section{Optimization framework}\label{sec:Method}

In this section, we first review the Wirtinger flow algorithm\cite{Phase_Wirtinger}, and then introduce the Wirtinger flow formulation into the FP reconstruction process, and incorporate the noise constraint. Finally, we derive the WFP reconstruction algorithm robust to the capturing noise.

\subsection{Review of the Wirtinger flow algorithm}

Wirtinger flow algorithm \cite{Phase_Wirtinger} is a recently reported technique to solve the standard phase retrieval problem. Specifically, it retrieves a plural signal ${\bf x}\in \mathbb{C}^{n}$ from a series of its real sampling measurements ${\bf b}\in \mathbb{R}^{m}$, with the measurement formation defined as ${\bf b} = |{\bf A}{\bf x}|^2=\bf (Ax)^*\odot Ax$. Here ${\bf A}\in \mathbb{C}^{m\times n}$ is a linear sampling matrix, and $\odot$ stands for the dot product.

Based on the quadratic loss function, the Wirtinger flow algorithm transforms the phase retrieval task into a minimization problem as
\begin{eqnarray}\label{eqs:Model_Ori}
\min && f({\bf x}) = \frac{1}{2}||{\bf (Ax)^*\odot Ax - b}||_F^2,~~~{\bf x}\in \mathbb{C}^n.
\end{eqnarray}
Here $||\cdot||_F$ is the Frobenius norm, and is calculated as $||\format X||_F = \sqrt{\sum_{i,j}\format X_{ij}^2}$.

Such an optimization model can be solved in an iterative manner, utilizing the gradient descent scheme. According to \cite{Wirtinger_Book_1,Wirtinger_Book_2}, the derivative of the complex quadratic cost function with respect to ${\bf x}^*$, i.e. $\frac{\partial f}{\partial {\bf x}^*}$, is necessary for updating $\bf x$ in each iteration. In implementation, $\bf x$ is updated in a gradient descending manner as \cite{Phase_Wirtinger}
\begin{eqnarray}\label{eqs:Update_Wirtinger}
{\bf x}^{(k+1)} = {\bf x}^{(k)} - \Delta\frac{\partial f}{\partial {\bf x}^*}|_{{\bf x} = {\bf x}^{(k)}}.
\end{eqnarray}
Here $\Delta$ is the gradient descent step size set by users, and $\frac{\partial f}{\partial {\bf x}^*}$ can be easily calculated according to the Wirtinger derivatives as
\begin{eqnarray}\label{eqs:Wirtinger_Gradient}
\frac{\partial f}{\partial {\bf x*}}
&=&  \frac{\partial \frac{1}{2}||{\bf (Ax)^*\odot Ax - b}||_F^2}{\partial {\bf x*}}\\\nonumber
&=& {\bf A}^H\left[(|{\bf Ax}|.^2 - {\bf b)\odot (Ax)}\right].
\end{eqnarray}
With the above derivations, the Wirtinger flow algorithm is summarized as Alg. \ref{alg:WirtingerFlow}.

\begin{algorithm}[ht]
\SetKwInOut{Majorization}{Majorization}\SetKwInOut{Minimization}{Minimization}
\SetKwData{set}{set}
\SetKwInOut{Initialization}{Initialization}\SetKwInOut{Input}{Input}\SetKwInOut{Output}{Output}
\vspace{2mm}
\Input{Sampling matrix $\bf A$, measurement vector $\format b$, initialization $\format x^{(0)}$.}
\Output{Retrieved plural signal $\bf x$.}
\vspace{2mm}
 $\format{x}=\format{x}^{(0)}$,
 $k = 0$\;
      \While{not converged}{
          Update $\format x^{(k+1)} $ according to Eq. (\ref{eqs:Update_Wirtinger}) and Eq. (\ref{eqs:Wirtinger_Gradient})\;
          $k:=k+1$.
          }
\caption{\small{Wirtinger flow algorithm}}
\label{alg:WirtingerFlow}
\end{algorithm}

\subsection{Wirtinger flow optimization for Fourier Ptychographic---WFP}
In terms of the FP reconstruction, the target is to recover the HR spatial spectrum from a series of LR images captured in spatial space. The relation between the HR reconstruction and the LR observations corresponds to two sequential linear operations: (i) down-sampling caused by the object aperture, and (ii) inverse Fourier transform to the LR spectrum bands caused by the microscope imaging system. We treat these two operations as a whole and use $\format A\in \mathbb{C}^{m\times n}$ to represent this combinational sampling process.
Specifically, denoting the HR spatial spectrum as a plural vector $\format x\in \mathbb{C}^n$, the corresponding sampling matrix $\format A = \format F\format S$ is composed of two components: inverse Fourier transform $\format F$ and down-sampling $\format S$.

In addition, we also consider the capturing noise explicitly, and the measurement formation model becomes
\begin{eqnarray}\label{eqs:Formation}
\format b = |\format A\format x|^2 + \format n,
\end{eqnarray}
where $\format n\in \mathbb{R}^{m}$ denotes the capturing noise, we assume which to be Gaussian. We use ${\sigma} \in \mathbb{R}$ to represent the standard deviation of the noise. The three sigma rule tells that, nearly all (99.73$\%$) the samples of a random variable lie within 3 times standard deviation from its mean. Therefore, we can approximatively formulate our noise constraint as
\begin{eqnarray}\label{eqs:Noise_Neq}
|\format n| \leqslant 3\sigma.
\end{eqnarray}
Introducing a relaxation vector $\bm \epsilon\in \mathbb{R}^{m}$, we can transform the above inequality into an equality
\begin{eqnarray}\label{eqs:Noise_Eq}
\format n\odot \format n - 9\sigma^2 + {\bm \epsilon} \odot {\bm \epsilon} = \bf 0.
\end{eqnarray}
Combining the above noise constraint (Eq. \ref{eqs:Noise_Eq}) with the measurement formation (Eq. \ref{eqs:Formation}), we can get the optimization model for FP reconstruction as
\begin{eqnarray}\label{eqs:Model}
\min && f({\bf x}) = \frac{1}{2}||{\bf (Ax)}^*\odot {\bf Ax} + {\bf n} - {\bf b}||_F^2 \\\nonumber
s.t. && \format n\odot \format n - 9\sigma^2 + {\bm \epsilon} \odot {\bm \epsilon} = \bf 0.
\end{eqnarray}

In the following, according to the Wirtinger flow algorithm, we derive the WFP optimization algorithm to solve the above model. First, we introduce a weighting parameter $\mu$ to incorporate the noise constraint into the objective function, and the model becomes
\begin{eqnarray}\label{eqs:ALM}
\min & f({\bf x}) = \frac{1}{2}||{\bf (Ax)}^*\odot {\bf Ax} + {\bf n} - {\bf b}||_F^2 + \frac{\mu}{2}||\format n\odot \format n - 9\sigma^2 + {\bm \epsilon} \odot {\bm \epsilon}||_F^2.
\end{eqnarray}
This is similar to the augmented Lagrangian function \cite{ALM_1}, which can be solved by sequentially updating each variable \cite{ALM_2}, while keeping the other variables constant. The updating can be conducted either by assigning zero to the function's partial derivative with respect to the updating variable, or by the gradient descent technique.
Here we utilize a similar scheme, and sequentially update the optimization variables, i.e., $\format x$, $\format n$ and $\bm \epsilon$, in Eq. \ref{eqs:ALM}.

For $\format x$, by calculating the partial derivative of $f$ to $\format x^*$, we can get its updating rule using the gradient descent technique as
\begin{eqnarray}\label{eqs:WFP_X}
{\bf x}^{(k+1)} &=& {\bf x}^{(k)} - \Delta_{{\bf x}}\frac{\partial f}{\partial {\bf x}^*}|_{{\bf x} = {\bf x}^{(k)}} \\\nonumber
&=& {\bf x}^{(k)} - \Delta_{{\bf x}}{\bf A}^H\left[(|{\bf Ax}|.^2 + {\bf n} - {\bf b)\odot (Ax)}\right]|_{{\bf x} = {\bf x}^{(k)}},
\end{eqnarray}
with $\Delta_{{\bf x}}$ being the gradient descent step size of $\bf x$.
Similarly, we can set the step size of $\format n$ as $\Delta_{{\bf n}}$, and update $\format n$ following
\begin{eqnarray}
\label{eqs:WFP_N}{\bf n}^{(k+1)} &=& {\bf n}^{(k)} - \Delta_{{\bf n}}\frac{\partial f}{\partial {\bf n}}|_{{\bf n} = {\bf n}^{(k)}} \\\nonumber
&=& {\bf n}^{(k)} - \Delta_{{\bf n}}\left[(|{\bf Az}|.^2 + {\bf n -b}) + \mu({\bf n\odot n} - 9{ \sigma}^2 + {\bm \epsilon\odot \bm \epsilon})\odot 2{\bf n}\right]|_{{\bf n} = {\bf n}^{(k)}}.
\end{eqnarray}

For updating of $\bm \epsilon$, we let the partial derivative of $f$ to $\bm \epsilon$ equal to $\bf 0$, and derive the closed-form updating rule as
\begin{eqnarray}\label{eqs:WFP_E}
& &\frac{\partial f}{\partial {\bm \epsilon}}|_{{\bm \epsilon} = {\bm \epsilon}^{(k+1)}} = \left[\mu({\bf n\odot n} - 9\sigma^2 + {\bm \epsilon\odot \bm \epsilon})\odot 2{\bm \epsilon}\right]|_{{\bm \epsilon} = {\bm \epsilon}^{(k+1)}}= {\format 0}\\\nonumber
&\Rightarrow& {\bm \epsilon}^{(k+1)} = \sqrt{\max\left(9\sigma^2 - {\bf n\odot n}, \format 0\right)}.
\end{eqnarray}


Based on the above derivations, the proposed WFP algorithm is summarized in Alg. \ref{alg:WFP}. As for the initialization, we set $\format x^{(0)}$ as the spatial spectrum of the up-sampled version of the LR image, which is captured under the normal incident light.


\begin{algorithm}[ht]
\SetKwInOut{Majorization}{Majorization}\SetKwInOut{Minimization}{Minimization}
\SetKwData{set}{set}
\SetKwInOut{Initialization}{Initialization}\SetKwInOut{Input}{Input}\SetKwInOut{Output}{Output}
\vspace{2mm}
\Input{Sampling matrix $\bf A$, measurement vector $\format b$, initialization $\format x^{(0)}$.}
\Output{Retrieved plural signal $\bf x$ (sample's HR spatial spectrum).}
\vspace{2mm}
 $\format{n}^{(0)}=\format{0}$,
 $\bm \epsilon^{(0)}=\format{0}$,
 $k = 0$\;
      \While{not converged}{
          Update $\format x^{(k+1)} $ according to Eq. (\ref{eqs:WFP_X})\;
          Update $\format n^{(k+1)} $ according to Eq. (\ref{eqs:WFP_N})\;
          Update $\bm \epsilon^{(k+1)} $ according to Eq. (\ref{eqs:WFP_E})\;
          $k:=k+1$.
          }
\caption{\small{WFP algorithm}}
\label{alg:WFP}
\end{algorithm}


As for the parameters settings of $\Delta_{{\bf x}}$ and $\Delta_{{\bf n}}$, similar to WF, we assign $\Delta_{{\bf x}}^{(k)} = \frac{\theta^{(k)}}{||\format x^{(0)}||^2}$ and $\Delta_{{\bf x}}^{(k)} = \frac{\theta^{(k)}}{||\bm \sigma||^2}$, where $\theta^{(k)} = \min\left(1-e^{-k/k_0},\theta_{max}\right)$. As stated in \cite{Phase_Wirtinger}, $k_0 = 330$ and $\theta_{max} = 0.4$ work well, so we also use these settings in our algorithm. We have released our
source code for non-commercial use, which can be downloaded \href{http://www.sites.google.com/site/lihengbian}{here}.

\section{Experiments}\label{sec:Experiments}

In this section, we conduct a series of experiments on both synthetic and real captured data to validate the proposed WFP algorithm.

\subsection{Experiment on synthetic data}\label{sec:Simulations}


\vspace{1mm}
\noindent\textbf{Algorithms for comparison:}\hspace{5mm}
To demonstrate the performance and advantages of the proposed WFP algorithm, here we run the WFP as well as the AP algorithm on simulated FP data. Besides, to investigate WFP's ability in addressing acquisition noise, we further compare its results with those produced by applying denoising before or after AP reconstruction.
Here we use BM3D\cite{BM3D} for denoising considering its promising results and high efficiency, as stated in \cite{BM3D_2}. For simplicity, we respectively refer to these two methods including the denoising operation as "BM3D+AP" and "AP+BM3D".

\vspace{1mm}
\noindent\textbf{Criterion:}\hspace{5mm}
Besides the visual results, we also utilize two quantitative criteria to assess the recovery performance of the above methods.  The first one is the peak signal to noise ratio (PSNR), which has traditionally been widely used to assess the quality of processed images compared to benchmark. PSNR intuitively describes the intensity difference between two images, and would be smaller for lower quality recovery.
Another adopted criterion is the structure similarity (SSIM)\cite{SSIM} that measures the spatial structural closeness between two images, and thus consists with human perception better than PSNR. The SSIM score ranges from 0 to 1, and is higher when two images are of more similar structural information.
Note that here both PSNR and SSIM are calculated on the intensity images.

\begin{figure*}[h]
\centering
\centerline{\includegraphics[width=1\textwidth]{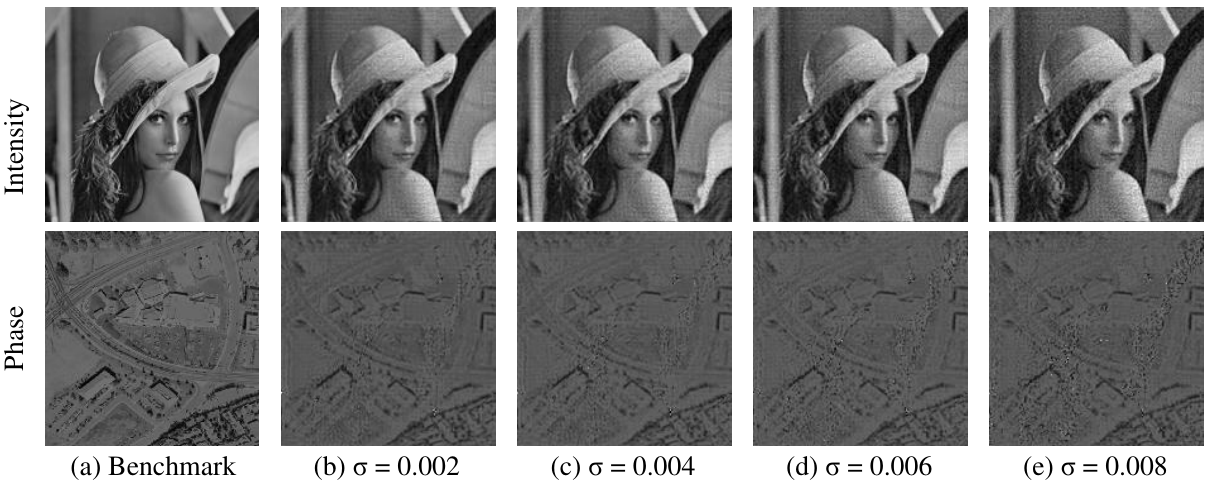}}
\vspace{-1mm}
\caption{Reconstruction results by WFP on different noise-level simulated data. The top row shows the intensity images, while the bottom row shows the phase images.}
\label{fig:Simulation_NoiseLevel}
\end{figure*}

\begin{table*}[h]
\centering
\caption{Quantitative comparison among the recovery of WFP at different noise levels.}\label{tab:Simulation_NoiseLevel}
{\small
\begin{tabular}{p{0.2\linewidth}<{\centering}p{0.15\linewidth}<{\centering}p{0.15\linewidth}<{\centering}p{0.15\linewidth}<{\centering}p{0.15\linewidth}<{\centering}}
\cline{1-5}
$\sigma$ & 0.002& 0.004 & 0.006 & 0.008 \\
\cline{1-5}
PSNR(dB) & 26.14    &  25.52    & 24.80 & 23.97\\
\cline{1-5}
SSIM     & 0.76   & 0.72  & 0.66  & 0.60\\
\cline{1-5}
\end{tabular}
}
\end{table*}

\vspace{1mm}
\noindent\textbf{Experiment parameter settings:}\hspace{5mm}
The convergence experiment in \cite{Phase_Wirtinger} shows that the Wirtinger flow algorithm works successfully when measurements are more than 6 times of the signal entries to be recovered. In terms of the FP problem, assuming that the overlap ratio is $\xi $, the LR images are of $m\times m$ pixels and we capture $k\times k$ LR images, the sampling ratio between measurements and signal entries can be calculated as
\begin{equation}\label{eqs:SamplingRatio}
\eta = \frac{m^2k^2}{\left[(1-\xi)m(k-1)+m\right]^2}
\approx \frac{1}{\left(1-\xi\right)^2}.
\end{equation}
Similar to \cite{FPM_Nature}, by setting $\xi = 65\%$, we can get the sampling ratio as around 8. The ratio is higher than the minimum convergence requisition ($\sim$6) in \cite{Phase_Wirtinger}. Therefore, we adopt the above experiment settings in our simulation experiment, namely $\xi = 65\%$, $k = 15$, $m = 100$.

\vspace{1mm}
\noindent\textbf{Results:}\hspace{5mm}
Based on the above specifications, the captured image volume in our simulation experiment is synthesized by following three steps: 1) we apply FFT to the original HR image, and select subregions corresponding to different incident angles, by multiplying the HR spectrum with an ideal pupil function (all ones in the pupil circle and zeros outside). 2) We shift these sub spatial spectra to the origin location, and do inverse Fourier transform to recover the LR plural images in the spatial domain. 3) We retain only the intensity of these LR plural images, and add Gaussian white noise to obtain the simulated captured noisy images. In our implementation, we use the 'Lena' and the 'Map' image ($512\times512$ pixels) from the USC-SIPI image database \cite{Data} as the HR intensity and phase image, respectively. The LR images' pixel numbers are set to be one tenth of the HR image along both directions.

First, we apply the proposed WFP on the simulated data with varying noise levels to study the algorithm's performance. Specifically, the standard derivation $\sigma$ of the additive noise ranges from 0.002 to 0.008 with a 0.002 interval. By algorithm testing, 500 iterations are enough for WFP to converge, and hence we set the iteration number of WFP as 500.
The visual and quantitative results are respectively shown in Fig. \ref{fig:Simulation_NoiseLevel} and Tab. \ref{tab:Simulation_NoiseLevel}.
From the results we can see that WPF works well to reconstruct both intensity and phase information. Besides, as the noise level increases, the reconstruction quality does not degenerate much. This illustrates the robustness of WFP to different noise levels, and thus a wider applicability.

\begin{figure*}[ht]
\centering
\centerline{\includegraphics[width=0.95\textwidth]{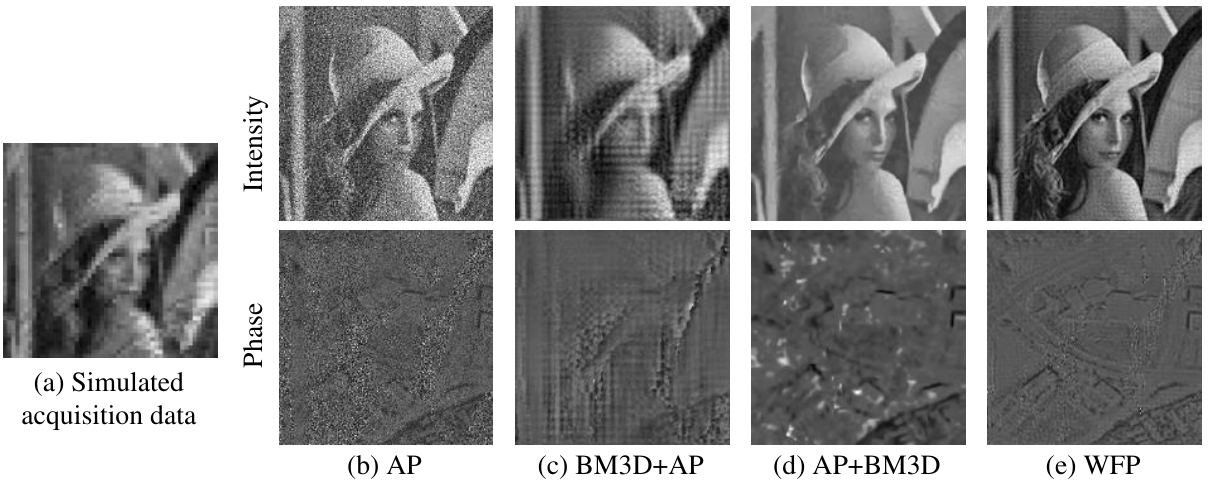}}
\vspace{-2mm}
\caption{Reconstruction results of different methods on the same dataset.}
\label{fig:Simulation_Methods}
\end{figure*}

Then, we compare WFP with the above mentioned three other methods, i.e., AP, BM3D+AP, and AP+BM3D, to show their pros and cons. Here the noise level is fixed at $\sigma = 0.004$. The iteration number of conventional AP is set to 50 to ensure convergence. The simulated acquisition image under the normal illumination is shown in Fig. \ref{fig:Simulation_Methods}(a). The quantitative and visual reconstruction results are respectively shown in Tab. \ref{tab:Simulation_Methods} and Fig. \ref{fig:Simulation_Methods}(b)-(d).

Due to the noise corruption, the SNR of captured images is very low, especially for the images corresponding to high spatial frequencies. As a result, the reconstruction intensity and phase image of AP in Fig. \ref{fig:Simulation_Methods}(b) are very noisy. When BM3D is applied before the AP reconstruction, a lot of high frequency information are filtered out, thus there exist serious artifacts in the final recovery, as shown in Fig. \ref{fig:Simulation_Methods}(c). If we apply BM3D after the AP reconstruction, though most of the noise is removed, many crucial image details are filtered out as well (see the areas of hat tassels in Fig .\ref{fig:Simulation_Methods}(d)).
Differently, the proposed WFP incorporates noise suppression into the reconstruction framework, and conducts these two operations jointly. This largely avoids error accumulation in successive processing and achieves higher performance.  
Consequently, WFP could obtain satisfying reconstruction results with less noise and more details.
In a nutshell, WFP largely outperforms the other three methods, on both visual and quantitative metrics.

\begin{table*}[ht]
\centering
\caption{Quantitative comparison among different reconstruction methods.}\label{tab:Simulation_Methods}
{\small
\begin{tabular}{p{0.2\linewidth}<{\centering}p{0.15\linewidth}<{\centering}p{0.15\linewidth}<{\centering}p{0.15\linewidth}<{\centering}p{0.15\linewidth}<{\centering}}
\cline{1-5}
& AP & BM3D+AP & AP+BM3D & WFP \\
\cline{1-5}
~PSNR(dB) & 13.33    &  20.03    & 18.71 & {\bf 25.52}\\
\cline{1-5}
~SSIM     & 0.18   & 0.41  & 0.66  & {\bf 0.72}\\
\cline{1-5}
~Running time &{\bf 12s} & 14s & 15s &  1min\\
\cline{1-5}
\end{tabular}
}
\end{table*}

The advantageous performance is at the expense of high computational cost. We implement all the four different methods using Matlab on an Intel i7 3.6 GHz CPU computer, with 16G RAM and 64 bit Windows 7 system. The running time of different methods are listed in the bottom row of  Tab.~\ref{tab:Simulation_Methods}. Obviously, WFP takes longer time than the other methods.

\subsection{Experiment on real captured data}

To further validate the effectiveness of WFP, we build a FPM setup to capture LR images as inputs for FP reconstruction. Similar to \cite{AFP},
an upright microscope with a 2$\times$ (NA = 0.1) objective is used in the platform. The LED array is placed around 8cm under the specimen, and the lateral distance between two adjacent LEDs is 4 mm. The central wavelength of incident light is 632nm. The pixel size of the captured raw images is $\sim$1.85$\mu$m.

\begin{figure*}[h]
\centering
\centerline{\includegraphics[width=\textwidth]{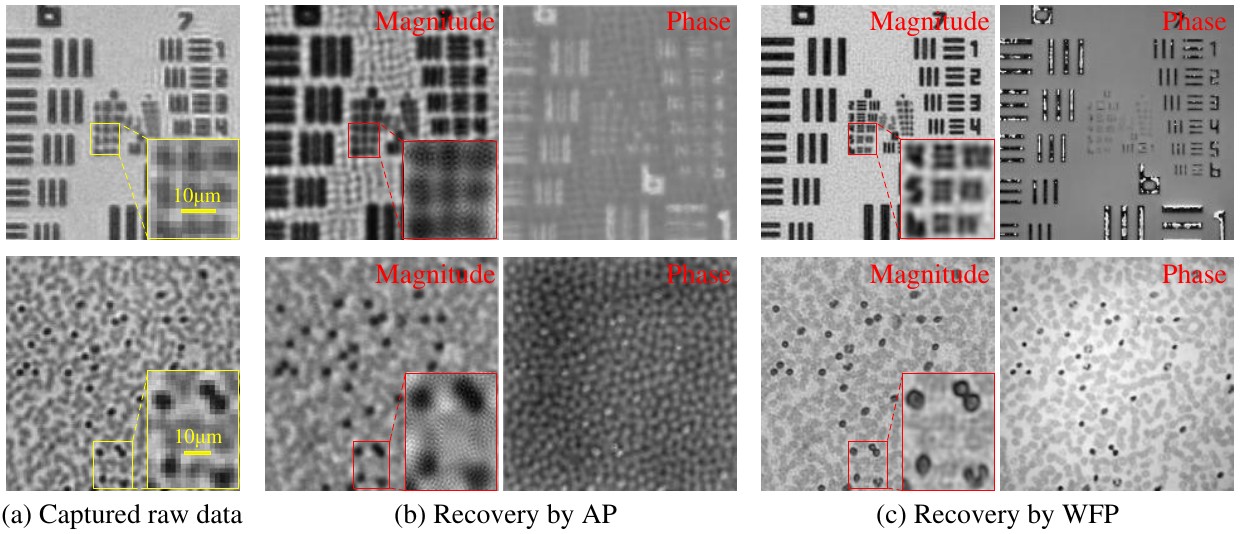}}
\vspace{-2mm}
\caption{Reconstruction comparison between AP and WFP on real captured data.}
\label{fig:Real_Experiment}
\end{figure*}

First, we capture the LR images corresponding to the 15$\times$15 LED positions, with the exposure time for each LED as 1ms, and apply AP as well as WFP to the image set for performance comprison. Fig.~\ref{fig:Real_Experiment}(a) shows the LR images of the USAF chart and the blood smear sample captured under the normal illumination.
The HR reconstruction results of AP and WFP are respectively presented in Fig. \ref{fig:Real_Experiment}(b) and Fig. \ref{fig:Real_Experiment}(c), where the left columns show recovered amplitudes, and the right columns present recovered phases.

From the results we can conclude three advantages of WFP over conventional AP. First, the reconstruction results of WFP own higher resolution than those of AP, and thus contain more details (see the close-ups for clearer comparison). Second, WFP could effectively suppress noise (see the smooth regions in the USAF chart). Third, WFP could reconstruct much more accurate phase than AP. Note that there exists some phase jumps recovered by WFP in the feature areas of the USAF chart, where the magnitudes are close to zero. This is due to that in these areas, the phase can be any value and their assignments would not affect successful magnitude recovery.

\begin{figure*}[ht]
\centering
\centerline{\includegraphics[width = 1\textwidth]{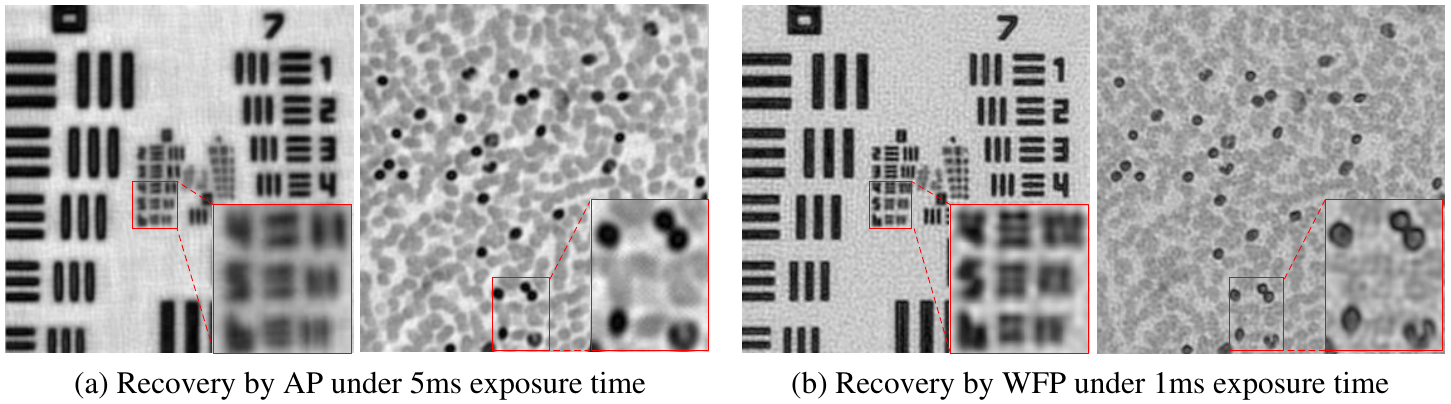}}
\vspace{-3mm}
\caption{Resolution demonstration of AP and WFP under different exposure time.}
\label{fig:TimeEvaluation}
\end{figure*}

Then, to quantitatively evaluate the advantage of WFP over conventional AP in the view of exposure time, we increase the exposure time, and apply AP to corresponding acquisition data under longer exposure time. See the amplitude reconstruction results in Fig. \ref{fig:TimeEvaluation}.
The proposed WFP could resolve a comparable resolution under 1ms exposure time, to that of the conventional AP under 5ms exposure time. This indicates that WFP could save round $80\%$ exposure time than AP to achieve the same reconstruction accuracy.

In all, WFP offers a feasible way for the FP technique to reconstruct highly accurate results when there exists non-ignorable capturing noise, such as the case of low exposure time, or when the hardware is not so precise. This will do lot of help in real applications.

\section{Conclusions and discussions}\label{sec:Conclusions}

This paper proposes a reconstruction framework termed Wirtinger flow optimization for Fourier Ptychography (WFP). Based on the recently reported Wirtinger flow algorithm, WFP formulates the FP recovery as a quadratic optimization problem, and presents a solution utilizing the gradient descent scheme. By incorporating priors on the capturing noise, WFP can save around 80$\%$ of the exposure time for the present FP technique, while without obvious performance degeneration.
Results on both synthetic and real captured data validate the effectiveness of WFP.

One extension of WFP is to handle non-uniform noise. This can be easily realized by treating the standard derivation of noise $\sigma$ in Eq. (\ref{eqs:Noise_Neq}) as spatial non-uniform, namely by changing ${\sigma} \in \mathbb{R}$ to ${\bm\sigma} \in \mathbb{R}^{m}$. Besides, as a flexible optimization framework, WFP can also be easily extended by introducing other priors and constraints. For example, we can incorporate the sparsity of the latent HR image \cite{Sparse_1} into our framework, which may further reduce the snapshot number and thus the acquisition time. We can also introduce the total variance prior \cite{TV_1, TV_2} into WFP to further suppress noise in the reconstruction results.
What's more, in the optimization model in Eqs. (\ref{eqs:Model}), the sampling matrix can be composed of any kinds of linear operations (down-sampling and inverse Fourier transform in conventional FP). Therefore, WFP is applicable for different variants of conventional FP, such as multiplexed FP \cite{FPM_Multiplexing_1,FPM_Multiplexing_2} and extended FP for fluorescence imaging \cite{FPM_Fluorescence}.

Although advantages in multiple ways over the conventional FP algorithm, WFP is limited in running efficiency, i.e., WFP needs more running time than AP.
Therefore, shortening the running time of WFP is one of our future work. Utilizing accelerated gradient descent methods and introducing parallel computation techniques are two promising speeding up options.

\end{document}